\documentclass[aps,prc,reprint,groupedaddress,amsmath,dviepsmx]{revtex4-1}

\usepackage{graphicx}
\usepackage{braket}
\usepackage{bm}
\usepackage{longtable}
\usepackage{color}

\usepackage{amsmath}	
\newcommand{\bvec}{\boldsymbol}

\newcommand{\OBe}{{}^{16}{\rm O}+{}^{8}{\rm Be}}
\newcommand{\CC}{{}^{12}{\rm C}+{}^{12}{\rm C}}
\newcommand{\Nea}{{}^{20}{\rm Ne}+\alpha}

\begin{document}

\title{Cluster correlation and nuclear vorticity  in low-lying $1^-$ states of $^{24}$Mg}

\author{Yohei~Chiba}
\affiliation{Department of Physics, Osaka City University, Osaka 558-8585, Japan}
\affiliation{Nambu Yoichiro Institute of Theoretical and Experimental Physics, Osaka City University, Osaka 558-8585, Japan}
\affiliation{Research Center for Nuclear Physics (RCNP), Osaka
University, Ibaraki 567-0047, Japan}
\author{Yoshiko Kanada-En'yo}
\affiliation{Department of Physics, Kyoto University, Kyoto 606-8502, Japan}
\author{Yuki Shikata}
\affiliation{Department of Physics, Kyoto University, Kyoto 606-8502, Japan}

\date{\today}

\begin{abstract}
\begin{description}
  \item[Background] Low-energy dipole states have been hot topics in stable and unstable nuclei. 
Recently, Nestrenko {\it et al.} proposed 
two low-energy dipole modes of the individual toroidal state and the compressional state in $^{24}$Mg. 
They associated the toroidal state with cluster structure, but there is no explicit analysis of the cluster structure. 
  \item[Purpose] Our purpose is to investigate low-lying $1^-$ states in $^{24}$Mg and clarify their natures
such as the dipole transition strengths, nuclear vorticity, and cluster features.
  \item[Method] 
Wave functions of  $1^-$ states of $^{24}$Mg are described with the antisymmetrized molecular dynamics framework
combined with the generator coordinate method. Excitation energies and dipole transition strengths are calculated. 
Cluster wave functions are explicitly taken into account  to reveal the role of  cluster correlations in $1^-$ states.
Intrinsic matter density and transition current density are analyzed.
  \item[Results]
Two low-lying dipole states, the $1^-_1$($K=1$) and $1^-_2$($K=0$), are obtained.
The $1^-_1$($K=1$) state has the
strongest isoscalar toroidal dipole strength and shows two vortexes structure in the intrinsic transition current density. 
The $1^-_2$($K=0$) state features the isoscalar 
compressional dipole strength and exhibits the $^{16}$O+$^{8}\textrm{Be}$ cluster correlation.
  \item[Conclusions] 
The toroidal and compressional dipole modes separately appear as the $K=1$ and $K=0$ states 
in the deformed $^{24}$Mg system. 
The $1^-(K=1)$ state is the toroidal dipole state with the strong nuclear vorticity but no prominent cluster structure, and 
the $1^-(K=0)$ state is 
the compressional dipole state having enhanced cluster structure but has the weaker vorticity.
\end{description}
\end{abstract}

\pacs{Valid PACS appear here}
\maketitle

\section{Introduction} \label{sec:1}

Isoscalar (IS) monopole and dipole excitations have been extensively 
investigated by $\alpha$ inelastic scattering experiments.
Significant low-energy IS strengths 
have been observed in various nuclei and attracting great interests (see for example,
Refs.~\cite{Harakeh-textbook,Paar:2007bk,Roca-Maza:2018ujj} and references therein).
A central issue is to reveal properties and origins of those low-energy dipole modes.

In order to understand the low-energy dipole modes, 
the vortical dipole mode (called also the torus or toroidal mode) has been originally 
proposed by hydrodynamical models \cite{semenko81,Ravenhall:1987thb}, and
later studied with  microscopic frameworks such as mean-field approaches
\cite{Paar:2007bk,Vretenar:2001te,Ryezayeva:2002zz,Papakonstantinou:2010ja,Kvasil:2011yk,Repko:2012rj,Kvasil:2013yca,Nesterenko:2016qiw,Nesterenko:2017rcc},
antisymmetrized molecular dynamics (AMD) \cite{Kanada-Enyo:2017fps,Kanada-Enyo:2017uzz,Kanada-Enyo:2019hrm}, and a cluster model \cite{Shikata:2019wdx}.
The vortical dipole mode is characterized by the vorticiy of the transition current and strongly excited by the 
toroidal dipole (TD) operator as discussed by Kvasil {\it et al.}\cite{Kvasil:2011yk}. These features are different from the  
compressional dipole (CD) mode which is excited by the standard IS dipole operator. 
Following Ref.~\cite{Kvasil:2011yk}, 
we call the vortical dipole mode ``TD mode''  to distinguish the compressional dipole (CD) mode.
The TD mode in deformed nuclei 
has been recently investigated in various stable and unstable nuclei in a wide mass-number region from light- to heavy-mass nuclei.
Cluster structures of the TD mode in  $p$-shell nuclei such as $^{12}$C and $^{10}$Be have been studied 
by the authors \cite{Kanada-Enyo:2017fps,Kanada-Enyo:2017uzz,Shikata:2019wdx}.

Very recently, Nesterenko {\it et al.} have investigated  dipole excitations in $^{24}$Mg with the 
Skyrme quasiparticle random-phase-approximation (QRPA) for axial-symmetric deformed nuclei,
and predicted that a TD state appears as the low-lying $K^\pi=1^-$ state \cite{Nesterenko:2017rcc}. 
In the nuclear current density of the TD mode, they found 
the vortex-antivortex type nuclear current in the deformed system and suggested its association with
the cluster structure of  $^{24}$Mg. 

For cluster structures of $^{24}$Mg, 
one of the authors and his collaborators have studied  positive-parity states of $^{24}$Mg
 with the AMD framework 
\cite{KanadaEnyo:1995tb,KanadaEnyo:1995ir,KanadaEn'yo:2001qw,KanadaEn'yo:2012bj},
and discussed roles of the cluster structures of $\Nea$, $\OBe$, and $\CC$ in the IS monopole excitations
\cite{Chiba:2015zxa}. 
Kimura {\it et al.} have investigated negative-parity states of $^{24}$Mg with the AMD, and discussed
triaxial deformations of the ground and negative-parity bands \cite{Kimura:2012-ptp}.

Our aim is to clarify natures of the low-lying dipole modes in $^{24}$Mg
such as vortical and cluster features as well as the IS dipole transition strengths. 
In order to describe dipole excitations,
we apply the 
constraint AMD method combined with the generator coordinate method (GCM). 
As for the constraint parameters for basis wave functions in the AMD+GCM, 
the quadrupole deformations ($\beta\gamma$)
\cite{Suhara:2009jb,Kimura:2012-ptp}
and the inter-cluster distance ($d$) \cite{Taniguchi:2004zz}
are adopted. 
This method is useful to analyze cluster correlations as well as intrinsic deformations because various cluster structures are explicitly 
taken into account in the $d$-constraint wave functions as proved in application to 
$^{28}$Si in Ref.~\cite{Chiba:2016zyz} which discussed the role of cluster structure in  IS monopole and dipole excitations.
In order to investigate properties of the low-lying $1^-$ states of $^{24}$Mg, 
the transition strengths are calculated for the TD and CD operators which can probe the 
vortical and compressional features, respectively. Nuclear vorticity is discussed in analysis of 
the intrinsic transition current density. Cluster correlations in the low-lying dipole excitations are also discussed. 

The paper is organized as follows. 
In Sect.~\ref{sec:2}, 
the framework of AMD+GCM with $\beta\gamma$- and $d$-constraints are explained.
Section  \ref{sec:3} shows 
the calculated results for basic properties of the dipole states, and 
Sect. \ref{sec:4} gives detailed analysis focusing on the vortical and cluster features.
Finally, the paper is summarized in section \ref{sec:5}. In appendix \ref{app:1}, 
definitions of the transition current density, dipole operators and transition strengths are explained. 

\section{Framework}  \label{sec:2}

 We briefly explain the present framework of the AMD+GCM method with the $\beta\gamma$- and $d$-constraints.
The  method is similar to that used in Ref.~\cite{Chiba:2016zyz}.
For the detail, the readers are directed to Refs.~\cite{KanadaEn'yo:2012bj,Suhara:2009jb,Kimura:2012-ptp,Taniguchi:2004zz,Chiba:2016zyz,Kimura:2003uf} and references therein.

\subsection{Hamiltonian and variational wave function}
The microscopic Hamiltonian for an $A$-nucleon system is given as  
\begin{align}
  {H} = \sum_{i}^{A}{{t}_i} - {t}_\textrm{c.m.} + \sum_{i<j}^{A}{{v}_{ij}^{NN}} +
 \sum_{i<j}^{A}{{v}_{ij}^\textrm{Coul}}.
  \label{hamiltonian}
\end{align}
Here, the first term is the kinetic energy, and  the center-of-mass kinetic energy 
${t}_\textrm{c.m.}$ is exactly subtracted.  As for the effective nuclear interaction $v^{NN}_{ij}$, 
we employ Gogny D1S interaction \cite{Berger:1991zza}. The Coulomb
interaction $v^\textrm{Coul}_{ij}$ is approximated by a sum of seven Gaussians.

The intrinsic wave function of AMD is given by a Slater determinant of single-nucleon wave functions
$\varphi_i$,
\begin{align}
 \Phi_\textrm{int} &= {\mathcal A}\left\{\varphi_1\varphi_2 \cdots \varphi_A \right\},\\
\varphi_i&= \phi_i({\bm r}) \chi_i \xi_i, \label{eq:singlewf}\\
 \phi_i({\bm r}) &= \exp\biggl\{-\sum_{\sigma=x,y,z}\nu_\sigma
 \Bigl(r_\sigma -\frac{Z_{i\sigma}}{\sqrt{\nu_\sigma}}\Bigr)^2\biggr\}, \\
 \chi_i &= a_i\chi_\uparrow + b_i\chi_\downarrow,
\end{align}
where $\chi_i$ is the spin part and $\xi_i$ is the isospin part fixed to be proton or neutron.
In the present version of AMD, the spaital part  $\phi_i({\bm r})$ is expressed by the deformed Gaussian wave packet centered
at $\bm Z_i$ with the width parameters $\nu_\sigma$ $(\sigma=x,y,z)$ which are common for all nucleons.

The Gaussian center parameter ($\bm{Z}_i$) and the nucleon-spin direction ($a_i$ and $b_i$) for each 
nucleon and the width parameters $\nu_\sigma$ are the variational parameters 
optimized by the energy variation \cite{Kimura:2003uf}.
The energy variation is performed for the parity-projected intrinsic wave function 
$\Phi^\pi=\frac{1+\pi{P}_r}{2}\Phi_\textrm{int}$ $(\pi=\pm)$. 

For the ground state, constraint of the quadrupole deformation ($\beta\gamma$-constraint) is imposed in the 
energy variation of the positive-parity wave function.
We use the parametrization $\beta$ and $\gamma$ of the triaxial deformation 
as described in Ref.~\cite{Kimura:2003uf} and get the $\beta\gamma$-deformed configuration for given 
$\beta$ and $\gamma$ values after the energy variation.
For $1^-$ states, the $\beta\gamma$-constraint energy variation is performed for the negative-parity wave function. 
In addition, cluster configurations are also obtained by 
constraint on the inter-cluster distance ($d$-constraint) in the 
energy variation of the negative-parity wave function, 
and they are combined with the $\beta\gamma$-deformed configurations. 
For the cluster configurations, we adopt {\it quasi-clusters}  proposed 
in Ref.~\cite{Taniguchi:2004zz}. Let us consider $C_1+C_2$ configuration consisting of two quasi-clusters $C_1$ and $C_2$ 
with the mass numbers $A_1$ and $A_2$ $(A_1+A_2=A)$, respectively.  Each quasi-cluster $C_j$ is defined as the group of $A_j$ nucleons, 
and the constraint is imposed on the inter-cluster  distance  $d_{A_1+A_2}$ between two quasi-clusters $C_1$ and $C_2$, which
is defined as 
\begin{align}
& d_{A_1+A_2} = |\bm R_{C_1} - \bm R_{C_2}|,\\
&(\bm R_{C_j})_\sigma = \frac{1}{A_j}\sum_{i\in C_j}\textrm{Re}\left[\frac{Z_{i\sigma}}{\sqrt{\nu_\sigma}}\right], 
\end{align}
where $\bm R_{C_j}$ is the center-of-mass position of the quasi-cluster $C_j$.
In the present work, we adopt the $\Nea$, $\OBe$, and $\CC$ configurations for $C_1+C_2$ of quasi-clusters. 

After the energy variation under each constraint of $(\beta,\gamma)$, $d_{20+4}$, $d_{18+8}$, and $d_{12+12}$, 
we obtain the basis wave functions optimized for various values of the constraint parameters,
and superpose them in the GCM calculation as explained later.
For simplicity, we number the obtained basis wave functions $\{\Phi^\pi(i)\}$ with the index $i$.

It should be stressed that the cluster wave function in the present framework is composed of 
not inert (frozen) clusters but quasi-clusters, which can contain cluster breaking effects such as the core polarization, dissociation, and excitation.
These effects are taken into account in the energy variation at a given value of the quasi-cluster distance $d_{A_1+A_2}$. 
Moreover, in the small distance limit, the cluster wave function
becomes equivalent to a deformed mean-field wave function because of the antisymmetrization of nucleons.
Along the distance parameter $d_{A_1+A_2}$, the $d$-constraint wave function describes the structure change from  
the one-center system of a mean-field configuration to the two-center system of the spatially developed $C_1+C_2$ clustering via intermediate configurations with cluster correlation (or formation) at the nuclear surface.

\subsection{Angular momentum projection and generator coordinate method}
After the energy variation with the constraints, the obtained basis wave functions are projected
to the angular momentum eigenstates, 
\begin{align}
 \Phi^{J\pi}_{MK}(i) = \frac{2J+1}{8\pi^2}\int d\Omega D^{J*}_{MK}(\Omega)R(\Omega)\Phi^\pi(i),
\end{align}
where $D^{J}_{MK}(\Omega)$ and $R(\Omega)$ are Wigner's $D$
function and the rotation operator, respectively. They are superposed to describe the final GCM wave function for 
the $J^\pi_{n}$ state, 
\begin{align}
 \Psi^{J\pi}_{M,n} = \sum_{K,i} c_n(K,i)  \Phi^{J\pi}_{MK}(i)\label{eq:gcmwf}. 
 \end{align}
Here the coefficients $c_n(K,i)$ are determined by diagonalization of the norm and Hamiltonian matrices
so as to satisfy Hill-Wheeler  (GCM) equation
\cite{Hill:1952jb,Griffin:1957zza}.

\section{Results}  \label{sec:3}
\subsection{Result of energy variation} \label{sec:3.1}

We describe properties of the $\beta\gamma$-deformed and cluster configurations obtained by the 
energy variation with the corresponding constraint. 

For the $\beta\gamma$-deformed configurations, we obtain
almost the same result as those in the previous AMD study \cite{Kimura:2012-ptp}. 
In  the $J^\pi$-projected energy surface on the $\beta$-$\gamma$ plane 
obtained from the  $\beta\gamma$-deformed configurations, 
we find the energy minimum state with triaxial deformation at ($\beta,\gamma)=(0.49,13^\circ)$ for  $J^\pi=0^+$
and that at  ($\beta,\gamma)=(0.5,25^\circ)$ for $J^\pi=1^-$.
These deformed states at the energy minimums become the dominant component of the $0^+_1$ and $1^-_1$ states in the 
final result of the GCM calculation. 

\begin{figure*}
\includegraphics[width=1.0\hsize]{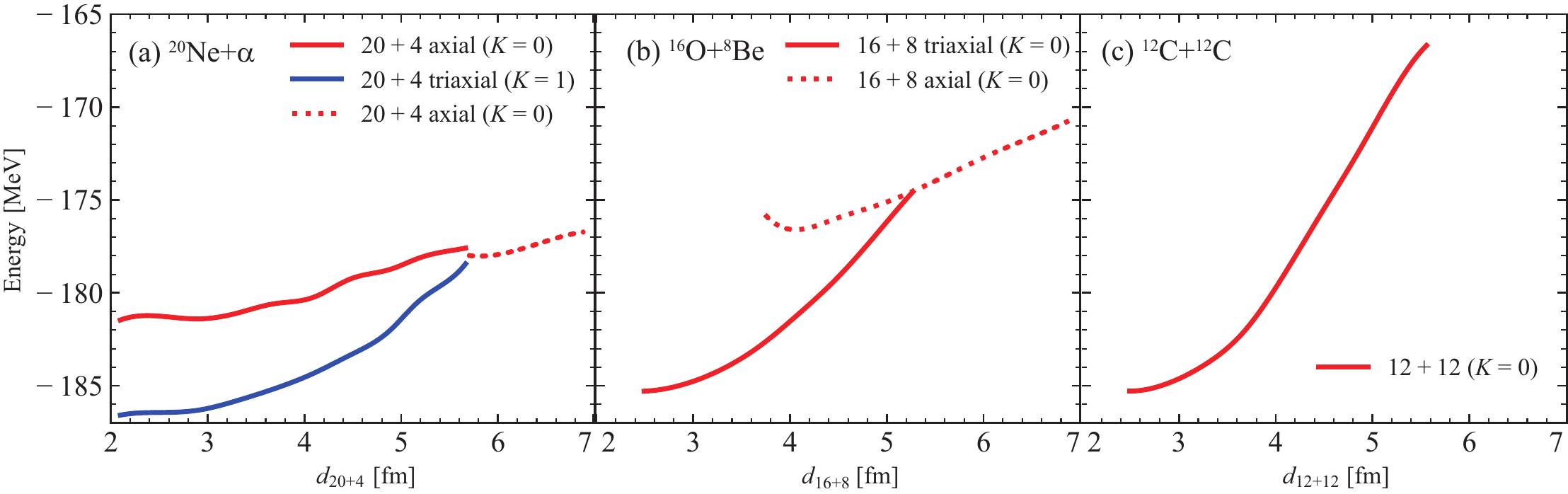}
\caption{Energy curves for $\Nea(20+4)$, $\OBe(16+8)$, and $\CC(12+12)$  quasi-cluster configurations
obtained by the 
$d$-constraint energy variation for negative parity states.
The $J^\pi=1^-$ projected energies are plotted as functions of the quasi-cluster distances $d_{A_1+A_2}$.
(a) $\Nea(20+4)$ configuration: energies of $K=0$ and $K=1$ states projected from triaxially deformed intrinsic states and 
$K=0$ states projected from axially deformed intrinsic states. 
(b) $\OBe(16+8)$ configuration:  energies of $K=0$ states projected from triaxially deformed intrinsic states and 
those projected from axially deformed intrinsic states. 
(c) $\CC(12+12)$ configuration: energies of $K=0$ states. 
}
  \label{fig:curve_d}
\end{figure*}

\begin{figure}
  \includegraphics[width=1.0\hsize]{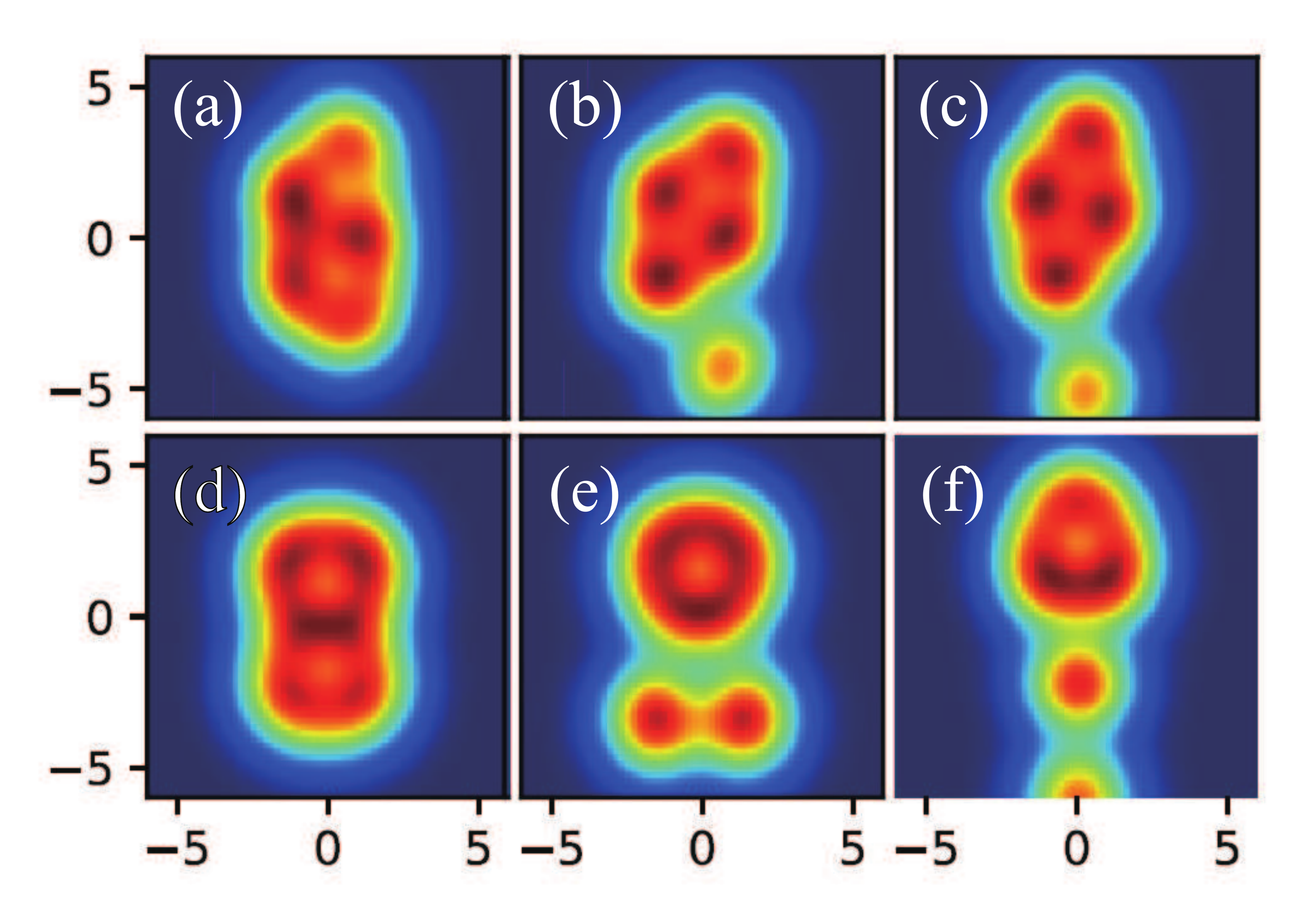}
 \caption{
(Color online) Intrinsic matter density distributions of
 the $\Nea$ and $\OBe$ quasi-cluster configurations obtained by the $d$-constraint energy variation for negative parity states. 
The panels (a), (b), and (c)  show the $\Nea$
configurations at $d_{20+4} =$ 2.5, 4.9, and 5.9 fm. 
The panels (d),  (e), and (f) show the
$\OBe$ configurations at  $d_{16+8} =$ 2.5, 4.9, and 6.1 fm.
The densities sliced at $x=0$ plane ($y-z$ plane) are shown. The units of the horizontal($y$) and vertical($z$) axes are fm.}
 \label{fig:dens}
\end{figure}

For the cluster configurations, we adopt the $\Nea$, $\OBe$,  and 
$\CC$ quasi-clusters as described previously. 
The calculated $J^\pi=1^-$ energies 
are shown as functions of quasi-cluster distances in Fig.~\ref{fig:curve_d}, and 
intrinsic density distributions are displayed in Fig.~\ref{fig:dens}. 

For the $\Nea$($20+4$) quasi-cluster configuration, the energy curves are shown in Fig.~\ref{fig:curve_d} (a) 
and the density distributions at $d_{20+4} =$ 2.5, 4.9 and 5.9 fm
are shown in the panels (a)$-$(c) of Fig.~\ref{fig:dens}. 
In the $2.0 \leq d_{20+4} \leq 5.7$ fm region, the triaxially deformed $\Nea$ configurations 
are obtained by the $d$-constraint energy variation and they 
yield the $K=0$ and $K=1$ states by the $J^\pi=1^-$ projection.
The $K=1$ energy curve is always lower than the $K=0$ energy curve.
The energy difference between the $K=1$ and $K=0$ states is about 6 MeV at $d_{20+4} = 2.0$ fm 
but it decreases to approximately 0 MeV at $d_{20+4}=5.7$ fm. In  $d_{20+4} \geq 5.7$ fm region, 
almost axial symmetric states 
with the dominant $K=0$ component are obtained for the $\Nea$ configuration
(see Fig.~\ref{fig:dens} (c) and the dotted line of Fig.~\ref{fig:curve_d} (a)). 

Figure \ref{fig:curve_d} (b) and Fig.~\ref{fig:dens}(d)$-$(f) show the energy curves and intrinsic density distributions 
for the $\OBe$($16+8$) quasi-cluster configuration. In $d_{16+8} < 5.4$ fm region, the lowest $\OBe$ configuration
is the triaxially deformed configuration because of two $\alpha$ clusters oriented along 
$y$ axis as shown in the intrinsic densities (d) and (e) of Fig.~\ref{fig:dens}.  
It contains 
only the  $K=\textrm{even}$ component because of the reflection symmetry with respect to $\pi$ rotation around $z$(longitudinal) axis. 
In $d_{16+8} \geq 5.4$ fm region, the axially symmetric $\OBe$
configuration (Fig.~\ref{fig:dens} (f)) becomes lowest as shown by the dotted line of Fig.~\ref{fig:curve_d} (b). 

In both cases of the $\Nea$ and $\OBe$ configurations, 
the intrinsic wave functions in the small quasi-cluster distance ($d_{A_1+A_2}$) region 
show no prominent cluster structure but have large overlap with the $\beta\gamma$-deformed configuration with triaxial
deformations. 
As the distance $d_{A_1+A_2}$ increases, the energies of the 
$^{20}$Ne+$\alpha$ and $^{16}$O+$^{8}\textrm{Be}$ configurations 
increase gradually indicating that the system is soft against spatial development of the cluster structures.
Compared with the $^{20}$Ne+$\alpha$ and $^{16}$O+$^{8}\textrm{Be}$ configurations, 
the energy of the $\CC$ configuration increases rapidly as $d_{12+12}$ increases as shown in Fig.~\ref{fig:curve_d} (c) 
because such the symmetric cluster configuration is relatively unfavored in the negative parity ($K^\pi=0^-$) state.
As a result, inclusion of $\CC$ cluster configurations gives 
almost no contribution to the low-lying $1^-$ states in the GCM calculation.

\subsection{GCM result of dipole excitations} \label{3.2}
We present the GCM result obtained using all 
the $\beta\gamma$-deformed and cluster configurations. We focus
on the low-lying $1^-$ states and their isoscalar dipole strengths. 
The definitions of the CD and TD operators and transition strengths 
are explained in appendix \ref{app:1}. 

\subsubsection{Spectra and transition strengths}
\begin{figure}
  \includegraphics[width=1.0\hsize]{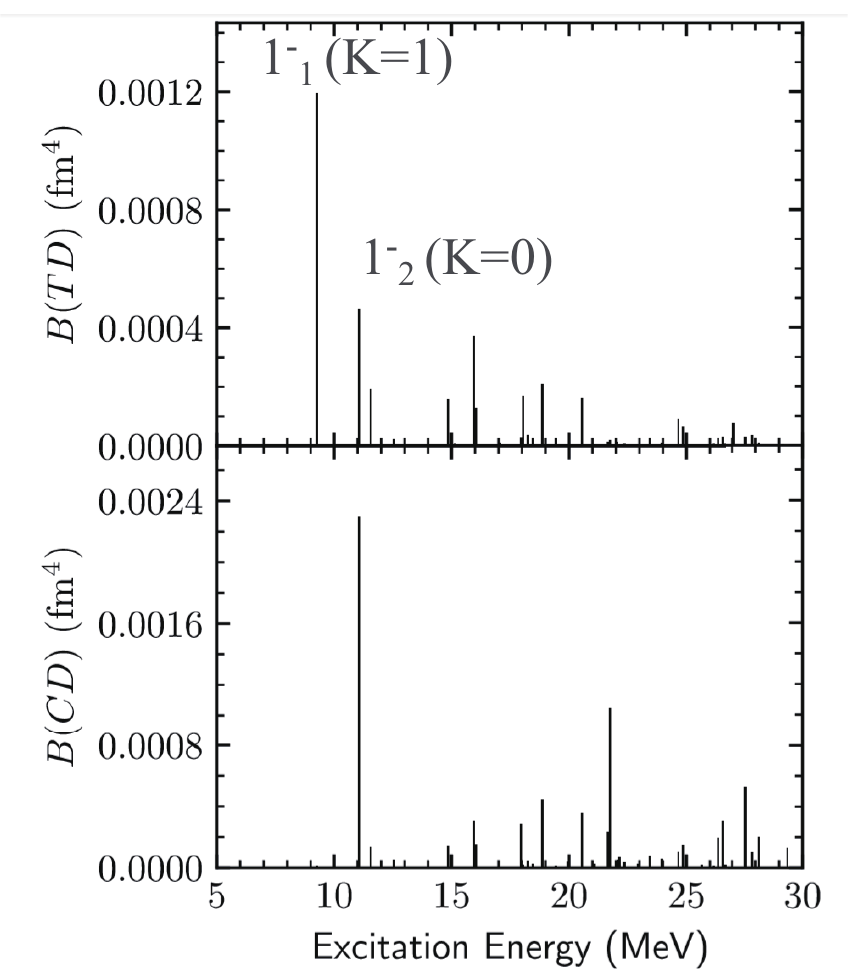}
	\caption{Strength functions of dipole transitions. 
The toroidal and compressional dipole strengths, $B(\textrm{TD})$ and $B(\textrm{CD})$,
	are plotted as functions of $1^-$ excitation energies  in the upper and bottom panels, respectively.}
 \label{fig:strength}
\end{figure}

The calculated CD and TD transition strengths ($B(\textrm{CD})$ and $B(\textrm{TD})$)
are plotted with respect to the $1^-$ excitation energies ($E_x$) in Fig.~\ref{fig:strength}.
In the low-energy region $E_x\approx 10$ MeV, 
we obtain  two dipole states, the $1^-_1$ and $1^-_2$ states, which have quite different natures from each other. 
One is the $1^-_1$ state at $E_x= 9.5$ MeV with the strongest  TD transition,  
and the other is the $1^-_2$ state at $E_x=11.2$ MeV with the significant CD transition strength.
Therefore, the $1^-_1$ state can be regarded as the TD mode, and the $1^-_2$ state is the low-lying CD mode.

In analysis of the dominant $K$ component in these two states, one can assign 
the former (TD mode) to the band-head state of a $K=1$ band and 
the latter (CD mode) to a $K=0$ band. This separation of the $K=1$ and $K=0$ components in the triaxially deformed intrinsic system
plays a key role in the low-lying TD and CD modes in $^{24}$Mg.
To emphasize this feature of $K$ quanta,  we denote the $1^-_1$  and $1^-_2$ states as $1^-_{K=1}$  and $1^-_{K=0}$, respectively, 
in the following.
The energy ordering of these two $1^-$ states in our result is consistent with 
the results of $\beta\gamma$-AMD \cite{Kimura:2012-ptp} and  QRPA \cite{Nesterenko:2017rcc} calculations. 
In the experimental data, $1^-$(7.56 MeV) and $1^-$(8.44 MeV) states were tentatively 
assigned to $K=0$ and $K=1$, respectively\cite{Fifield:1979gfv}.
Therefore, the theoretical  $1^-_{K=1}$ and  $1^-_{K=0}$ states in the present result may correspond
to the experimental $1^-$(8.44 MeV) and $1^-$(7.56 MeV)  states though the energy 
ordering of two states seems inconsistent with the observation.

\subsubsection{Cluster correlations}
We here discuss roles of cluster correlations in the $1^-_{K=1}$ and $1^-_{K=0}$ states.
To discuss the cluster correlation effect, 
we perform the GCM calculation 
using only the $\beta\gamma$-configurations but without the cluster configurations 
and compare the results with and without cluster configurations. 
The excitation energies and transition strengths of the $1^-_{K=1}$ and $1^-_{K=0}$ states
calculated with and without cluster configurations are
summarized in Table \ref{tab:cc}.

The correlation energies induced by the cluster correlations can be evaluated by the 
energy gain by inclusion of the cluster configurations. 
The energy gain is 0.3 MeV for the 
$1^-_{K=1}$ state and 1.0 MeV for the $1^-_{K=0}$ state.
The large energy gain in the $1^-_{K=0}$  state 
indicates significant cluster correlation, which mainly comes 
from the $\OBe$ configuration. 
The cluster correlation from the $\OBe$ configuration
 also contributes to the CD transition strength of 
the $1^-_{K=0}$ state as 50\% enhancement of $B(\textrm{CD})$.
This result is understood by the general feature that 
the low-energy ISD strengths can be enhanced by  asymmetric clustering as discussed in 
Ref.~\cite{Chiba:2015khu}.
Compared with the $1^-_{K=0}$ state, 
the properties of the $1^-_{K=1}$ state is not affected so much by inclusion of cluster configurations. 

\begin{table}[h]
   \caption{The calculated values of excitation energies ($E_x$) and TD and CD strengths 
of the $1^-_{K=1}$ and $1^-_{K=0}$ states 
obtained by the GCM calculations with and without the cluster configurations (cc). 
}
   \label{tab:cc}
   \begin{ruledtabular}
   \begin{tabular}{ccccc}
    &
    \multicolumn{2}{c}{$K=1$ state}
    &
    \multicolumn{2}{c}{$K=0$ state} \\
          & w/ cc & w/o cc & w/ cc & w/o cc \\
          \hline
   $E_x$ (MeV)                       & 9.52 & 9.85 & 11.21 & 12.18   \\
   $B(\textrm{TD})$ ($10^{-3} {\rm fm}^4$)  & 1.20 &  1.13   & 0.41   & 0.32      \\
   $B(\textrm{CD})$ ($10^{-3} {\rm fm}^4 $)  & 0.00 &   0.00   & 2.38  & 1.61    \\
   \end{tabular}
   \end{ruledtabular}
\end{table}

\section{Discussions}\label{sec:4}

\subsection{Cluster correlations in $1^-_{K=1}$ and $1^-_{K=0}$}

In the previous discussion, we showed that inclusion of cluster configurations gives
significant contributions to the  $1^-_{K=0}$ state but relatively minor effect on the $1^-_{K=1}$ state.  
However, it does not necessarily mean no cluster correlation in the $1^-_{K=1}$ state
because $\beta\gamma$-deformed configurations can implicitly contain cluster correlations.
What we have shown  
in the previous analysis of change by inclusion of the cluster configurations is just the effects from prominent cluster structures, 
which are beyond the $\beta\gamma$-constraint method.

For more detailed investigation of  cluster components in the $1^-_{K=1}$ and $1^-_{K=0}$ states, 
we calculate overlap of the GCM wave function with each basis of quasi-cluster configurations. 
The $1^-_{K=1}$ state has 89 $\%$ overlap with the $\Nea$ configuration at $d_{20+4}=2.5$ fm
projected to $J^\pi=1^-(K=1)$, which indicates significant $\Nea$ component.
Similarly, the $1^-_{K=0}$ state is dominantly described by $K=0$ component  of the $\OBe$ configuration at $d_{16+8}=2.5$ fm 
with $88\%$  overlap. 
The $1^-_{K=0}$ state also  has non negligible
overlap with spatially developed $\OBe$ configurations, e.g., 23\% overlap at $d_{16+8}=4.9$ fm. These
developed $\OBe$ cluster components
contribute to enhancement of the CD transition strength discussed previously. 

\subsection{Vorticity of the nuclear current}

\begin{figure}[t!]
  \includegraphics[width=\hsize]{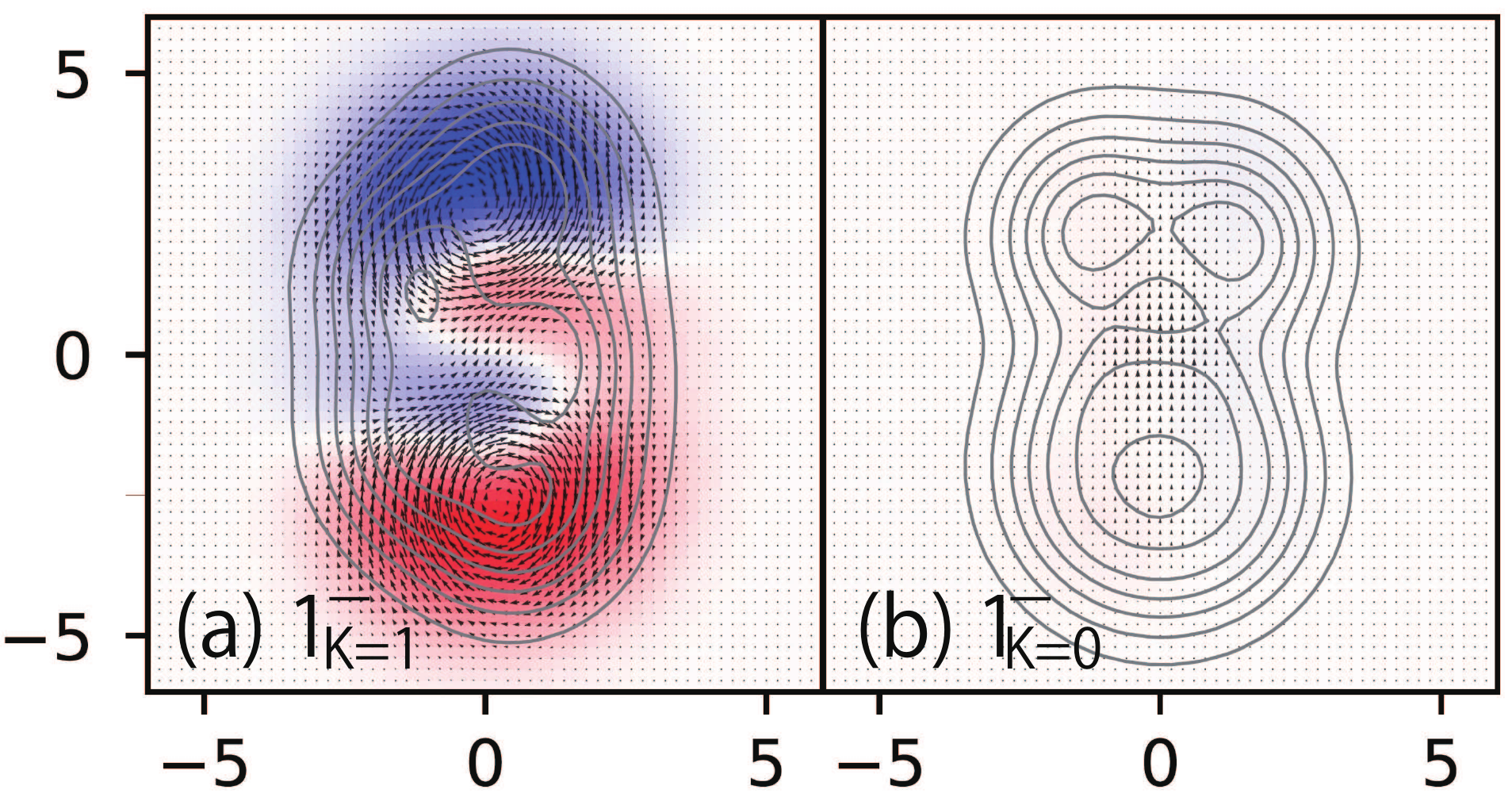}
  \caption{
The intrinsic transition current density $\delta \bm{j}(\bm{r})$ after the parity projection 
for $0^+_1\to 1^-_{K=1}$ and $0^+_1\to 1^-_{K=0}$ transitions.
The arrows and color plots indicate $\delta \bm{j}(\bm{r})$ and $x$-component of the vorticity $\bm{\nabla} \times \delta \bm{j}(\bm{r})$.
The contours are intrinsic matter densities of the $1^-_{K=1}$ and $1^-_{K=0}$ states before the parity projection. 
The densities sliced at $x=0$ plane ($y-z$ plane) are shown. The units of the horizontal($y$) and vertical($z$) axes are fm.}
	\label{fig:current}
\end{figure}

In order to reveal vortical nature of the two low-energy dipole modes, 
we analyze the intrinsic transition current density $\delta\bm{j}(\bm{r})$
of $0^+_1\to1^-_{K=1}$ and $0^+_1\to 1^-_{K=0}$ transitions. 
For simplicity, we take the dominant configuration of each state as an approximate
intrinsic state, and compute the transition current density in the intrinsic frame:
we choose the $\beta\gamma$-deformed configuration at  $(\beta,\gamma)=(0.49,13^\circ)$
for the ground state, 
the $\Nea$ configuration at $d_{20+4} = 2.5$ fm for the $1^-_{K=1}$ state, and 
the $\OBe$ configuration at $d_{16+8} = 2.5$ fm for the $1^-_{K=0}$.
In Fig.~\ref{fig:current}, 
the transition current density $\delta \bvec{j}$ and  vorticity $\nabla \times \delta \bvec{j}$ calculated 
after the parity projection are displayed 
by vector and color plots, respectively. The intrinsic 
matter density distribution of the $1^-$ states before the parity projection is also shown
by contour plot. 
 
In the transition current density in the $1^-_{K=1}$ excitation (Fig.~\ref{fig:current}(a)), 
one can see two vortexes  with opposite directions in the upper and lower parts of the longitudinal matter density.
The opposite vorticity is a specific character of the vortical dipole mode with $K=1$ 
in an elongated deformation, and consistent with the dipole mode called the 
vortex-antivortex configuration in Ref.~\cite{Nesterenko:2017rcc}. 
On the other hand, the transition current density in the  $1^-_{K=1}$ excitation (Fig.~\ref{fig:current}(b)) 
shows no vortex but irrotational flow with compressional nature along the $z$-axis (the longitudinal direction), 
which contributes to the CD dipole strength.
The difference in the vortical nature between the $1^-_{K=0}$ and $1^-_{K=1}$ states can be
more clearly seen in color plots of the vorticity.  
The $1^-_{K=1}$  excitation indicates the strong nuclear vorticity in the top and bottom
edge parts of the elongated shape, but the $1^-_{K=0}$  excitation shows much weaker  vorticity. 

\subsection{Cluster and single-particle natures of $1^-_{K=1}$ state} 

As described previously, 
the $1^-_{K=1}$ state is approximately described by the $^{20}$Ne+$\alpha$ cluster configuration 
at $d_{20+4}=2.5$ fm, which does not show a spatially developed clustering but the cluster correlation in the triaxially deformed state. 
As can be seen in the intrinsic density distribution shown in Fig.~\ref{fig:dens}(a), 
the essential cluster correlation in the $1^-_{K=1}$ state
is formation of $\alpha$ clusters caused by four nucleon correlations  at the nuclear surface. 
In a schematic picture, the cluster correlation in the $1^-_{K=1}$ state is associated with 
the $^{16}$O core with two $\alpha$ clusters in one side of the core. 
Two $\alpha$ clusters are placed at the surface of the $^{16}$O core in a tilted configuration 
and yields the $K=1$ component because of the asymmetry against the $\pi$ rotation around $z$ axis.
On the other hand, the ground state has the triaxial deformation because of the $2\alpha$ correlation 
aligned in a normal direction along the surface of $^{16}$O. 
Then, the dipole excitation from the ground state to the  $1^-_{K=1}$ state 
can be understood by the vibrational (tilting) motion of the $2\alpha$ orientation 
at the surface of the $^{16}$O core. This tilting motion of the  $2\alpha$ clustering  produces the nuclear vorticity. 
Then, the vortex is duplicated in both sides because of the antisymmetrization effect and parity projection. 

It is also worth of mentioning a link between cluster and mean-field pictures for the 
$1^-_{K=1}$ mode by considering the small limit of the inter-cluster distance, where
the cluster structure can be associated with a deformed harmonic oscillator configuration. We here use 
the representation  $(n_xn_yn_z)$ with oscillator quanta $n_\sigma$ in $\sigma$ axis 
for a single-particle orbit in the deformed harmonic oscillator.  
In this limit,  the ground state corresponds to the $(011)^4(002)^4$ configuration with triaxial deformation,
while the $1^-_{K=1}$ state is regarded as $(011)^3(002)^4(003)^1$. It means that 
the $1^-_{K=1}$ transition is described by one-particle one-hole excitation of 
$(011)^{-1}(003)^1$ on the triaxially deformed ground state, which
induces the vortical nuclear current and contributes to the TD strength. 
This mechanism is similar to that discussed 
with the deformed mean-field approach in Ref.~\cite{Nesterenko:2017rcc}. However, we should remark 
that the present $1^-_{K=1}$ mode contains the cluster correlation
and corresponds to the coherent one-particle one-hole excitations in the $LS$-couping scheme. The 
coherent contribution from four nucleons in the SU(4) symmetry (spin-isospin symmetry) 
enhances collectivity of the vortical dipole excitation further than the $jj$-coupling configuration.

\section{Summary}\label{sec:5}

We investigated the low-lying $1^-$ states of $^{24}$Mg with the AMD+GCM framework 
with the $\beta\gamma$-constraint for the quadrupole deformation and the 
$d$-constraint for the $\Nea$, $\OBe$, and $\CC$ configurations. 
We discussed properties of the $1^-$ states such as IS dipole transition strengths, cluster correlations, and vortical nature.
In the low-energy region $E_x\approx 10$ MeV, we obtained the 
$1^-_{K=1}$  and $1^-_{K=0}$  states, which shows quite different features from each other. 
The $1^-_{K=1}$ is the toroidal dipole mode, 
which is characterized by the nuclear vorticity. 
The $1^-_{K=0}$  state has the significant compressional dipole strength
and the weaker vorticity. 
Effects of the cluster correlations on the excitation energy and transition strength of these two low-lying dipole states 
were analyzed. It was found that the spatially developed cluster configurations 
give significant contribution to the $1^-_{K=0}$ state, 
whereas the effect on the $1^-_{K=1}$ state is minor. 
We should stress that the deformation and cluster correlations play important roles in the low-energy dipole modes of
$^{24}$Mg.

\begin{acknowledgments}
The authors thank to Prof.~Nesterenko and Prof.~Kimura for fruitful discussions.
A part of the numerical calculations of this work were performed by using the
supercomputer at Yukawa Institute for theoretical physics, Kyoto University. This work was supported by 
MEXT/JSPS KAKENHI (Grant Nos. 16J03654, 18K03617,  18H05407, 18J20926, 18H05863, 19K21046).
\end{acknowledgments}

\appendix

\section{Definition of transition densities and dipole strengths}\label{app:1}

The density and current density operators for the nuclear matter are given as  
\begin{eqnarray}
\rho(\bvec{r})&=& \sum_k  \delta(\bvec{r}-\bvec{r}_k),\\
\bvec{j}(\bvec{r})&=& -\frac{i\hbar}{2m} \sum_k  \nabla_k\delta(\bvec{r}-\bvec{r}_k)+\delta(\bvec{r}-\bvec{r}_k)\nabla_k.
\end{eqnarray}  
For the current density, we consider only the convection term of the nuclear current but not the spin term of magnetization. 
The transition current density for the transition from the ground ($0^+_1$) to the $1^-$ states
is written as $\delta\bvec{j}(\bvec{r})= \langle 1^- |\bvec{j}(\bvec{r}) |0^+_1 \rangle$.  

For the dipole transition strengths, 
the following CD and TD operators are adopted,
\begin{eqnarray}
M_\textrm{CD}(\mu)&=&\frac{-i}{2\sqrt{3}c}\int d\bvec{r} \bvec{j}(\bvec{r}) 
 \nonumber\\
&\cdot& 
\left [  \frac{2\sqrt{2}}{5} r^2 \bvec{Y}_{12\mu}(\hat{\bvec{r}}) - r^2 \bvec{Y}_{10\mu} (\hat{\bvec{r}}) 
\right ],\\
M_\textrm{TD}(\mu)&=&\frac{-i}{2\sqrt{3}c}\int d\bvec{r} \bvec{j}(\bvec{r}) \nonumber\\
&\cdot& 
\left [
\frac{\sqrt{2}}{5} r^2 
\bvec{Y}_{12\mu}(\hat{\bvec{r}})+
r^2 \bvec{Y}_{10\mu} (\hat{\bvec{r}})  
\right ],
\end{eqnarray}
where $\bvec{Y}_{\lambda L\mu}$ is the vector spherical  harmonics. 
The CD operator corresponds to the standard IS dipole operator 
and is sensitive to the compressional dipole excitations, 
and the TD operator has been adopted in Ref.~\cite{Kvasil:2011yk} as a measure probing the nuclear dipole vorticity.
The CD and TD transition strengths are given by the square of the 
the reduced matrix elements of the corresponding dipole operators as 
$B(\textrm{CD,TD}; 0^+_1 \to 1^-)=\left| \langle 1^-||M_\textrm{CD,TD} ||0^+_1 \rangle \right |^2$. 
Using the continuity equation, the CD transition strengths is
related to the transition strength for the standard IS dipole 
operator $M_\textrm{IS1}(\mu)=\int d\bvec{r} \rho(\bvec{r}) r^3 Y_{1\mu} (\hat{\bvec{r}})$ as 
\begin{align}
B(\textrm{IS1})\equiv \left|\langle 1^-||M_\textrm{IS1} ||0^+_1 \rangle \right |^2
=\left(\frac{10\hbar c}{E}\right )^2 B(\textrm{CD}).
\end{align}


\begin{thebibliography}{9}
\bibitem{Harakeh-textbook}
M.N.~Harakeh, A.~van der Woude, Giant Resonances, Oxford University Press, 2001.

\bibitem{Paar:2007bk} 
  N.~Paar, D.~Vretenar, E.~Khan and G.~Colo,
  Rept.\ Prog.\ Phys.\  {\bf 70}, 691 (2007).

\bibitem{Roca-Maza:2018ujj} 
  X.~Roca-Maza and N.~Paar,
  Prog.\ Part.\ Nucl.\ Phys.\  {\bf 101}, 96 (2018)
  doi:10.1016/j.ppnp.2018.04.001
  [arXiv:1804.06256 [nucl-th]].


\bibitem{semenko81}
S. F. Semenko,  Sov. J. Nucl. Phys. {\bf 34}, 356 (1981).

\bibitem{Ravenhall:1987thb} 
  D.~G.~Ravenhall and J.~Wambach,
  Nucl.\ Phys.\ A {\bf 475}, 468 (1987).


\bibitem{Vretenar:2001te} 
  D.~Vretenar, N.~Paar, P.~Ring, and T. Nik\u{s}i\'{c}
  Phys.\ Rev.\ C {\bf 65}, 021301(R) (2002).


\bibitem{Ryezayeva:2002zz} 
  N.~Ryezayeva {\it et al.},
  Phys.\ Rev.\ Lett.\  {\bf 89}, 272502 (2002).


\bibitem{Papakonstantinou:2010ja} 
  P.~Papakonstantinou, V.~Y.~Ponomarev, R.~Roth and J.~Wambach,
  Eur.\ Phys.\ J.\ A {\bf 47}, 14 (2011).

\bibitem{Kvasil:2011yk} 
  J.~Kvasil, V.~O.~Nesterenko, W.~Kleinig, P.-G.~Reinhard and P.~Vesely,
  Phys.\ Rev.\ C {\bf 84}, 034303 (2011).

\bibitem{Repko:2012rj} 
  A.~Repko, P.-G.~Reinhard, V.~O.~Nesterenko and J.~Kvasil,
  Phys.\ Rev.\ C {\bf 87},  024305 (2013).

\bibitem{Kvasil:2013yca} 
  J.~Kvasil, V.~O.~Nesterenko, W.~Kleinig and P.-G.~Reinhard,
  Phys.\ Scripta {\bf 89}, 054023 (2014).

\bibitem{Nesterenko:2016qiw} 
  V.~O.~Nesterenko, J.~Kvasil, A.~Repko, W.~Kleinig and P.\ -G.~Reinhard,
  Phys.\ Atom.\ Nucl.\  {\bf 79}, 842 (2016).

\bibitem{Nesterenko:2017rcc} 
  V.~O.~Nesterenko, A.~Repko, J.~Kvasil and P.~G.~Reinhard,
  Phys.\ Rev.\ Lett.\  {\bf 120}, no. 18, 182501 (2018).

\bibitem{Kanada-Enyo:2017fps} 
  Y.~Kanada-En'yo, Y.~Shikata and H.~Morita,
  Phys.\ Rev.\ C {\bf 97}, no. 1, 014303 (2018).
\bibitem{Kanada-Enyo:2017uzz} 
  Y.~Kanada-En'yo and Y.~Shikata,
  Phys.\ Rev.\ C {\bf 95}, no. 6, 064319 (2017).

\bibitem{Shikata:2019wdx} 
  Y.~Shikata, Y.~Kanada-En'yo and H.~Morita,
  arXiv:1902.10962 [nucl-th].

\bibitem{Kanada-Enyo:2019hrm} 
  Y.~Kanada-En'yo and Y.~Shikata,
  Phys.\ Rev.\ C {\bf 100}, no. 1, 014301 (2019)
  doi:10.1103/PhysRevC.100.014301
  [arXiv:1903.01075 [nucl-th]].

\bibitem{KanadaEnyo:1995tb}
  Y.~Kanada-En'yo, H.~Horiuchi and A.~Ono,
  Phys.\ Rev.\  C {\bf 52}, 628  (1995).

\bibitem{KanadaEnyo:1995ir}
  Y.~Kanada-En'yo and H.~Horiuchi,
  Phys.\ Rev.\  C {\bf 52}, 647 (1995).

\bibitem{KanadaEn'yo:2001qw} 
  Y.~Kanada-En'yo and H.~Horiuchi,
  Prog.\ Theor.\ Phys.\ Suppl.\  {\bf 142}, 205 (2001).

\bibitem{KanadaEn'yo:2012bj}
  Y.~Kanada-En'yo, M.~Kimura and A.~Ono,
  PTEP {\bf 2012},  01A202 (2012).

\bibitem{Kimura:2012-ptp} 
M. Kimura, R. Yoshida, and M. Isaka,  Prog.\ Theor.\ Phys.\ {\bf 127}, 287 (2012).

\bibitem{Chiba:2015zxa} 
  Y.~Chiba and M.~Kimura,
  Phys.\ Rev.\ C {\bf 91}, no. 6, 061302(R) (2015)
  doi:10.1103/PhysRevC.91.061302
  [arXiv:1502.06325 [nucl-th]].

\bibitem{Suhara:2009jb} 
  T.~Suhara and Y.~Kanada-En'yo,
  Prog.\ Theor.\ Phys.\  {\bf 123}, 303 (2010)
  doi:10.1143/PTP.123.303
  [arXiv:0909.2218 [nucl-th]].

\bibitem{Taniguchi:2004zz} 
  Y.~Taniguchi, M.~Kimura and H.~Horiuchi,
  Prog.\ Theor.\ Phys.\  {\bf 112}, 475 (2004)
  doi:10.1143/PTP.112.475
  [nucl-th/0405014].

\bibitem{Chiba:2016zyz} 
  Y.~Chiba, Y.~Taniguchi and M.~Kimura,
  Phys.\ Rev.\ C {\bf 95}, no. 4, 044328 (2017)
  doi:10.1103/PhysRevC.95.044328
  [arXiv:1610.04000 [nucl-th]].
\bibitem{Kimura:2003uf} 
  M.~Kimura,
  Phys.\ Rev.\ C {\bf 69}, 044319 (2004)
  doi:10.1103/PhysRevC.69.044319
  [nucl-th/0311062].

\bibitem{Berger:1991zza} 
  J.~F.~Berger, M.~Girod and D.~Gogny,
  Comput.\ Phys.\ Commun.\  {\bf 63}, 365 (1991).
  doi:10.1016/0010-4655(91)90263-K

\bibitem{Hill:1952jb} 
  D.~L.~Hill and J.~A.~Wheeler,
  Phys.\ Rev.\  {\bf 89}, 1102 (1953).
  doi:10.1103/PhysRev.89.1102

\bibitem{Griffin:1957zza} 
  J.~J.~Griffin and J.~A.~Wheeler,
  Phys.\ Rev.\  {\bf 108}, 311 (1957).
  doi:10.1103/PhysRev.108.311


\bibitem{Fifield:1979gfv} 
  L.~K.~Fifield, E.~F.~Garman, M.~J.~Hurst, T.~J.~M.~Symons, F.~Watt, C.~H.~Zimmerman and K.~W.~Allen,
  Nucl.\ Phys.\ A {\bf 322}, 1 (1979).
  doi:10.1016/0375-9474(79)90329-4

\bibitem{Chiba:2015khu} 
  Y.~Chiba, M.~Kimura and Y.~Taniguchi,
  Phys.\ Rev.\ C {\bf 93}, no. 3, 034319 (2016)
  doi:10.1103/PhysRevC.93.034319
  [arXiv:1512.08214 [nucl-th]].


\end{thebibliography}
\end{document}